# Detection of slow atoms confined in a Cesium vapor cell by spatially separated pump and probe laser beams

Petko TODOROV[1*], Nikolay PETROV[1], Isabelle MAURIN[2],
Solomon SALTIEL[**][2,3], Daniel BLOCH[2]

[1] Institute of Electronics –BAS, Sofia, Bulgaria
[2] Laboratoire de Physique des Lasers, Université Paris 13, Sorbonne Paris-Cité,
UMR 7538 du CNRS, 99 Avenue J-.B. Clément, F-93430 Villetaneuse, France
[3] Physics Department, Sofia University, 5, J. Bourchier Boulevard, 1164 Sofia, Bulgaria

## ABSTRACT

The velocity distribution of atoms in a thermal gas is usually described through a Maxwell-Boltzman distribution of energy, and assumes isotropy. As a consequence, the probability for an atom to leave the surface under an azimuth angle θ should evolve as cos θ, in spite of the fact that there is no microscopic basis to justify such a law. The contribution of atoms moving at a grazing incidence towards or from the surface, *i.e.* atoms with a small normal velocity, here called "slow" atoms, reveals essential in the development of spectroscopic methods probing a dilute atomic vapor in the vicinity of a surface, enabling a sub-Doppler resolution under a normal incidence irradiation. The probability for such "slow" atoms may be reduced by surface roughness and atom-surface interaction. Here, we describe a method to observe and to count these slow atoms relying on a mechanical discrimination, through spatially separated pump and probe beams. We also report on our experimental progresses toward such a goal.

## 1. ATOMIC FLIGHT AT A GRAZING INCIDENCE WITH RESPECT TO A SURFACE

The kinetic theory of gas is a solid basis for thermodynamics, so that when considering a gas at thermal equilibrium, it is natural to assume that the atomic velocities are distributed according to a Maxwellian distribution of kinetic energy, and also that this vector distribution is isotropic. These two points may be not perfectly equivalent, as the anisotropic shape of the container may be susceptible to induce anisotropy in the vector distribution. Subtle discussions, notably around the consequences of the so-called "detailed balance" (between incoming, and outcoming atoms, relatively to the surface of the gas container) sometimes appear. The kinetic theory only considers ideal containers, *i.e.* ideal surfaces at the boundary of the gas region. It is by a flux equilibrium between incoming and departing atoms to the surface that one derives the Knudsen law for a rarefied gas ("molecular regime") of a "cos θ " probability of direction for a departing atom, with θ the angle of the departing atom trajectory relatively to the normal to the surface. Such a "cos θ" law is well-known, but has no connection with a microscopic description (for a discussion on these topics, see *e.g.* the review by Comsa and David [1]). It cannot apply for accommodation effects (surface and gas at a different temperature as often occurs for aeronautics studies), nor it can stand for a detailed study of atomic desorption from a well-characterized surface; nevertheless it remains widely accepted when various averagings have to be introduced in the modelling. The tradition for using such a "cos θ " model is so high that detailed desorption studies performed under high vacuum, rather than in a thermal gas surrounding, often describe the angular behavior along a "cos θ " expansion, *i.e.* a f(θ) law as f(θ) = $\Sigma_n a_n (\cos\theta)^n$ (see [1], and for examples of high value of the exponent n, see *e.g.* [2]).

---

[*] e-mail: petkoatodorov@yahoo.com
[**] now deceased

Experimentally, it is uneasy to perform the corresponding measurements, which demands a gas density low enough to make observable genuine surface effects. The high sensitivity of laser spectroscopy, and its ability to resolve different atomic velocities, makes it an attractive tool for experimental tests of this law. At least two such experiments have found agreement with the "cos θ " law, one [3] with a thermal gas cell (with not much details about the nature of the surface), the other one [4] more connected with desorption processes for a specific surface.

In all cases, the angular dependence has not been specifically tested for atoms departing from the surface under a grazing incidence. Such a situation of grazing incidence appears to be of a specific interest. Indeed, there is nowadays a sustained activity of laser spectroscopy for a gas confined close to a surface [5-15], either through the observation of reflection [5] (corresponding to an optical confinement, on a depth on the order of one reduced wavelength), either through confinement in a thin vapor cell [6-15] (confinement effects can appear for a thickness as large as 1 mm if the vapor is dilute enough [6], thicknesses below 100 nm are achievable [15]). Even for a single beam irradiation under normal incidence and a micrometric thickness[6-8], the observed spectroscopic signals can exhibit a contribution insensitive to the Doppler broadening associated to thermal motion. This provides a reason to consider these techniques as possibly competing with well-established techniques such as the (nonlinear) technique of (velocity-selective) saturated absorption in a gas cell, or even with linear sub-Doppler spectroscopy in a molecular beam. For these techniques relying on the confinement of the vapor, the contribution of atoms with velocities nearly parallel to the wall, *i.e.* atoms leaving the surface under a grazing incidence, is enhanced and can even be made totally dominant, notably in non linear spectroscopic schemes [6,8-14], or when a FM technique is coupled to linear spectroscopy [7]. Relatively to the normal velocity component, those atoms with flight nearly parallel to the wall are "slow atoms". Practically, the corresponding experiments have been so far performed only with alkali metal resonances, and the relative velocity selection is typically governed by the ratio of the homogenous width ($\geq$ 5 MHz natural width for Cs) to the characteristic Doppler broadening (~200 MHz for Cs). This yields a velocity selection not better than ~ 5m/s (to be compared with a thermal velocity ~ 200 m/s for Cs), already corresponding to an angular selection θ $\geq$ 88-89°. In the principle, one could extend these methods to narrower transitions such as an alkali-earth transition - see *e.g.* some calculations in [10] - or to some molecular transitions, at the extent of a tighter velocity-selection. In the standard technique of saturated absorption in a macroscopic gas cell, a velocity selection as narrow as $10^{-3}$, or even better, can currently be operated on narrow transitions (*i.e.* selected velocity at the level of a fraction of m/s). Conversely, for a confinement close to a surface, one can suspect that the roughness of the surface on the one hand, and the van der Waals attraction exerted by the surface on the other hand, limit the existence of such "slow" atoms, notably when assuming linear atomic trajectories. This makes it interesting to test experimentally the effective density of these grazing incidence angle atoms, providing the motivation for our work. We describe below the principle of our method for such a measurement, based upon a probing of atoms in a free flight after having been pumped in a different location, and we report on our experimental progress toward such a measurement.

## 2. PRINCIPLE FOR AN EXPERIMENTAL MEASUREMENT

In a dilute vapor cell of a micrometric thickness, irradiated under normal incidence by a single beam, the optical pumping to a non absorbing level, such as a ground state hyperfine level of an alkali vapor (*e.g.* the clock transition of Cs at 9.192 GHz), is governed by its transient evolution [6,8]: only slow atoms reach the steady state (the medium is no longer absorbing), and the optical pumping is hence a velocity-selective process. The major assumption is that atom trajectories are wall-to-wall, as atom-atom collisions are neglected owing to the dilute nature of the vapor. In the principle[6,8], the thicker the cell, the narrower the velocity selection can be, provided the vapor remains sufficiently dilute. We have already described in details[8] the effectiveness of this velocity selection. However, the natural width of the transition makes uneasy the optical observation of atoms slower than the "natural" velocity selection $v_{nat}$, defined by $v_{nat} = u_{th} \gamma_{nat} / \Gamma_{Dopp}$ with $\gamma_{nat}$ the natural width of the transition, $\Gamma_{Dopp}$ its Doppler width, and $u_{th}$ the thermal velocity. To observe these slow atoms, one can rather consider a mechanical velocity selection, assuming a known (and constant) thickness L, within which atoms "marked" by a pump beam must propagate on a distance R in order to be probed[9]. Typically, the normal velocity that can be detected in such a scheme is v ~ L/R $u_{th}$, allowing a large experimental flexibility if L remains micrometric (*e.g.* 10-100 μm), while R is a macroscopic distance (mm or even cm range) that must remain compatible with the expected mean free path. A preliminary experiment[9] on this principle, with two spatially separated beams – of a quasi-Gaussian shape- had shown the possibility to observe in Cs vapor a contribution of atoms with a normal velocity roughly estimated to be < 2 m/s. No systematic measurements, nor theory, had been reported on this topic.

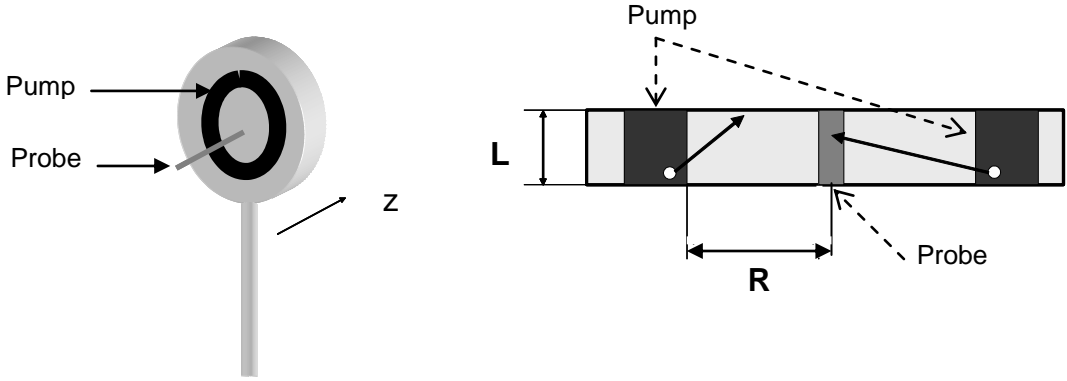

Fig. 1: *Principle of the experiment; a thin cell – thickness L, much smaller than the overall diameter of the cell- is irradiated by a ring-shaped pump beam. Only atoms with a slow normal velocity can travel from the pump region to the axial probe region.*

Here, we consider a situation in which the pump beam is ring-shaped, and the probe beam, sent under a normal incidence, is just located on the centre of the pump beam ring. The interest of such a configuration is that it maximizes the signal, as all atoms going through the probe beam with sufficiently slow velocities will have been pumped. Although a larger experimental flexibility can be offered if two different lasers are used, there is no restriction in using the same laser for the pump and probe beams. The probe beam intensity should be non saturating, as the principle of the quantitative evaluation of the number of slow atoms reaching the probe beam after having been pumped, lies in the measurement of the absorption change as induced by the pump. To count the pumped atoms travelling across the probe beam, it is simple to assume that the probe absorption is operated in its steady-state regime: practically, the transient build-up of absorption can be neglected, especially for the slow atoms which are here of interest as long as the cell thickness largely exceeds a wavelength [7]. When scanning the frequency of the (normal incidence) probe beam, the atoms pumped at a distance appear only as a Doppler-free resonance. For a better evaluation of the distance travelled by the detected atoms, the diameter of the probe beam should be as small as possible (up to the limit of a focusing that would lead to a saturating intensity). Conversely, the pump beam should provide an efficient transfer (to the other hyperfine ground state level, or more generally to the other branch of a $\Lambda$-type system), and must be strongly saturating: for an alkali-metal resonance, it is easy to reach these strongly saturating intensities, owing to the very long relaxation time constants of the optical pumping process, as long as the pumping is operated on a non-cycling transition. This standard hypothesis of a negligible relaxation of the optical pumping makes the external diameter of the ring-shaped pump a non relevant parameter; rather, the shape of the pump-ring, on its internal diameter, must be well defined (*cf.* section 3, and fig. 6) in order to avoid optical pumping for atoms located in the dark region separating the pump and probe beams (this may occur, even at a reduced rate, in the presence of a residual intensity in the "dark" region)

In an independent work[11], a calculation of the lineshape has been performed as function of the L and R parameters, under the usual assumption that the velocity distribution in the thin cell obeys a 3D-Maxwellian distribution, actually described in cylindrical coordinates as the product of a Gaussian distribution for $v_z$, as given by $f(v_z) = (u_{th}\pi^{1/2})^{-1} \exp - (v_z^2/u_{th}^2)$, by an isotropic 2D Gaussian distribution for $r = (x^2+ y^2)^{1/2}$. Here, we concentrate on the idea that some slow atoms may be missing, and we analyze the influence of a missing "slice" in the velocity distribution of slow atoms. Figure 2 shows the relative velocity distribution of pumped atoms in the situations R/L = 10, and R/L = 30. These two situations do not correspond to a sharp velocity selectivity: in an experiment with a thin cell of a current thickness ($\leq$ 100 μm), this corresponds to a small (sub-mm) dark region. The peculiarity of the geometry imposes a sharp contribution for the distribution of slow atoms; also, the tails (respectively above 25 m/s, and 8.3 m/s) correspond to atoms which are fast enough– *i.e.* 3-D velocity exceeding $u_{th}$ - to travel under the required grazing incidence. From fig.2, one easily deduces that the probe spectrum is expected to be Doppler-free, or at least to be largely sub-Doppler if the R/L ratio remains small. Assuming a modified velocity distribution $f(v_z) = 0$ for $|v_z| \leq v_{missing}$, $f(v_z) = (u_{th}\pi^{1/2})^{-1} \exp - (v_z^2/u_{th}^2)$ for $|v_z| > v_{missing}$, in which too "slow" (normal velocity) atoms are missing, one easily derives (see fig.3) the distribution of the (normal) velocity of pumped atoms travelling to the probe region. Figure 4 shows that when varying R/L (*i.e.* by changing the pump beam size), the amplitude of the probe absorption, as measured on resonance, yields a sensitive indication of the missing velocities. Although inverting the details of a velocity distribution smoother than the abrupt one considered presently would be a hard task, the reasons evoked in section 1 about the possible absence of slow atoms probably justify that the effective distribution could be rather abrupt.

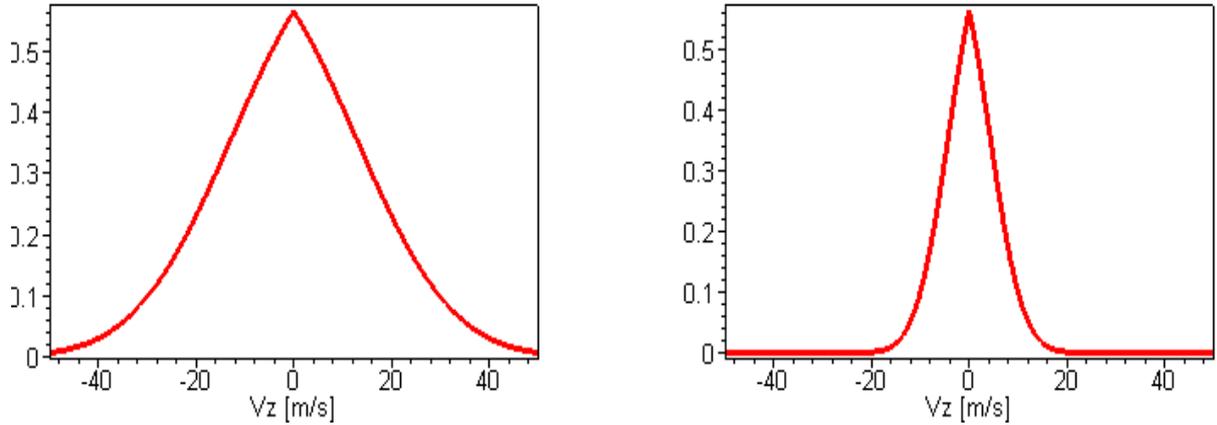

Fig.2  Velocity distribution of atoms reaching the probe region, after having travelled from the pump region. A Maxwellian distribution is assumed, with a thermal velocity $u_{th}$ = 250 m/s. The R/L ratio is 10 (left) or 30 (right). The pump intensity is assumed to be so strong that the calculation does not depend on the pump frequency detuning.

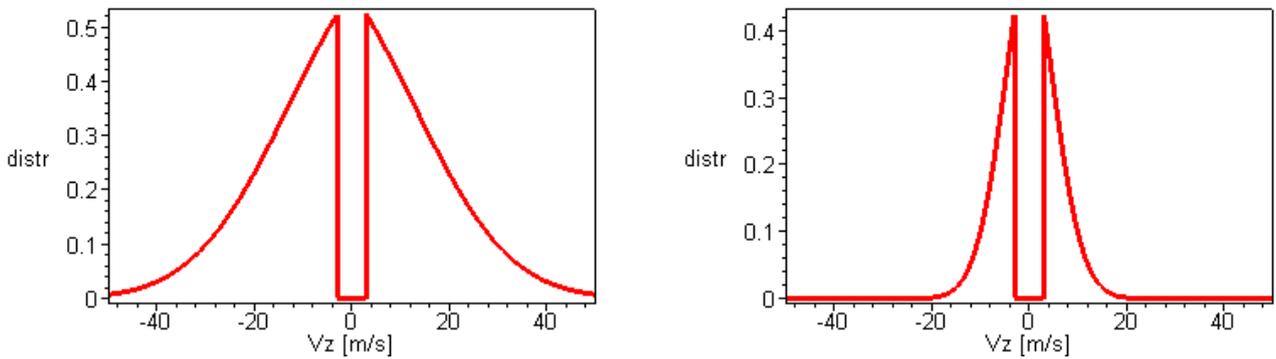

Fig.3  Same as fig. 2, assuming an initial velocity distribution for which $|v_z| < 3$ m/s atoms do not exist.

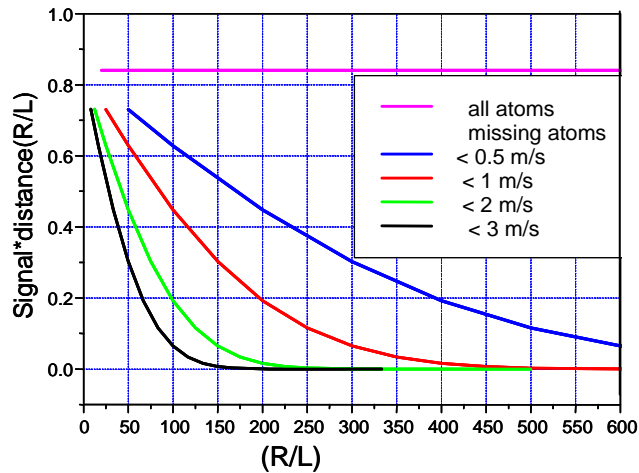

Fig. 4 : Amplitude of the probe absorption as a function of R/L, for various distributions of (normal) atomic velocities. The height is here normalized, so that for a distribution with no loss of slow atoms, the expected reduction of signal when increasing the pump-probe separation is already taken into account. One has taken $u_{th}$ = 250 m/s.

To implement the corresponding experiment, whose goal would be to check the experimental curve equivalent to fig. 4, two major initial assumptions are worth to be emphasized:
(i) Atom-atom collisions should remain negligible. Although this is the basic condition for a vapor to be considered as dilute in a thin cell, this condition may be more demanding (in the sense of requesting a low-pressure) when precisely looking for the "slowest" atoms : this is because long wall-to-wall trajectories allow for more time for a collision. This tends to impose that the experiments are performed under various temperatures, or gas pressures, in order to look for the extrapolation at a low pressure. Also, it makes it critical to eliminate residual buffer gases in the cell.
(ii) It is known that the optical pumping of an alkali atom can survive many collisions (more than 1000 ! ) with some special coatings (paraffin, silane, ...) to the wall. In such a situation, pumped atoms can be seen nearly everywhere, owing to their very poor relaxation. For an ordinary surface (*i.e.* glass), it is expected that hyperfine relaxation occurs on the wall. However, it is not established that a full loss of memory occurs on every single wall collision. Previous experiments[16] have suggested that a non negligible fraction of hyperfine optical pumping could survive an elementary wall collision. This may be highly problematic for our method. However, as long as the rate of this wall relaxation relaxation is not negligible (a problem to be addressed experimentally on the very same surface as the one considered), a "slow" atom with a nearly parallel flight will be mostly scattered under a standard direction, *i.e.* it will no longer be a slow atom, and the chances to reach the probe region after a wall collision will remain weak. Moreover, the mean dwell time on the surface [16] may lead to a temporal behavior allowing one to discriminate between direct flights, and flights to the probe region with a stop onto a surface. To solve this problem of atom-surface collision, it is conceivable to implement a "thin cell" in which part of the dark region between the pump and the probe beam would be of a much thicker width. This would be at the expense of a greater technical difficulty for the realization of the cell, but the atom-surface collisions, leading to a re-entry of pumped atoms, would be avoided in this zone.

## 3. EXPERIMENTAL SET-UP and PRELIMINARY EXPERIMENTS

We are presently implementing the corresponding experiments on Cs vapor confined in a thin cell with sapphire windows, with the spectroscopic measurements to be performed on the Cs $D_2$ line. Our cell has been specially designed, and realized in the D. Sarkisyan group in Armenia. It is essentially a cylindrical thin cell, with a diameter ~ 3 cm, with a Cs deposited in a bottom reservoir (as schematized in fig.1). The windows are made of a standard high quality polished sapphire. Sapphire can currently be of a small roughness, which is important for an attempt of a specific detection of atoms leaving the surface at a grazing incidence, and presently the roughness has been estimated to ~ 5 nm. The planarity and parallelism of the windows are also essential parameters for the accuracy of our experiment. We had measured interferometrically the thickness to be ~19 μm. When performing this thickness measurement in various points, we find the deviation to parallelism not to exceed 100 nm over 1 mm (*i.e.* parallelism as good as $10^{-4}$ rad). The cell (and the Cs reservoir) is generally heated up to ~60-70 °C to increase the Cs atomic density. From our experience with comparable cells, nor the thickness, nor the parallelism should be affected by a moderate temperature change. The typical "slow atoms" (for the normal velocity) on the $D_2$ line for a standard sub-Doppler experiment are atoms with $v_z \leq 5$ m/s, corresponding to a rather small pump-probe separation of 1 mm. This makes the design of this cell well suited for measurements of velocities in the 1-5 m/s range, at the possible expense of a low signal. Thicker cells, of a typical thickness 50 μm and 100 μm, and with a similar design, are expected to be produced for comparison, and also in view of a check of the velocity distribution on larger ranges, for which there are stronger reasons to expect a standard Maxwellian distribution.

For the spectroscopic measurements, we use a single laser, tunable around the $D_2$ line (852 nm). It is a narrow linewidth (< 3MHz) DFB type semiconductor laser, delivering a relatively high power (150 mW) sufficient to provide an intense (highly saturating) pump beam, even for a spatially extended beam. Apart from an auxiliary fraction sent to a saturated absorption set-up, simply providing a frequency reference, the laser beam is split into a standard weak probe beam, and into a pump beam, modulated by a mechanical chopper (typically at 2 kHz) and shaped into a diverging ring by optical elements (see fig. 5). The cell is small enough so that the pump beam divergence is negligible over the cell thickness. The probe transmission is detected through a lock-in detection, synchronous with the chopping of the pump. To better eliminate spurious residual modulated scattering, the pump and probe beam propagate in opposite directions. The most essential point in the set-up is hence the shaping of the pump beam, with the aim to convert as much power as possible into a ring-shape geometry. For this purpose, and following a spatial filtering of the pump beam, we have implemented a special transmission grating with circular grooves, which converts the beam into (conically) diverging diffraction rings. An axicon lens allows a reduction of the diverging rings (half-angle is ~ 11° for the first order). To suppress the zeroth diffracted order, a central mask is put on the pump beam, close to the window; in addition, a mask is also directly located on the central part of the axicon. The weaker higher order diffraction rings of the pump beam are simply eliminated by the limited aperture of our optics, and cannot have any influence on the signal. An example of the resulting spatial distribution of intensity for the pump beam is shown in fig. 6. The divergence of the ring-shaped pump beam allows a convenient way to vary the spatial pump-probe spatial separation by moving the position of the Cs cell. For an easier detection of the probe beam (with no loss induced by a beam splitter), a small angle is introduced between the probe beam and the pump axis, with the probe beam remaining at normal incidence. This small angle does not affect the pump beam efficiency as long as the residual Doppler broadening remains negligible relatively to the power-broadened transition width.

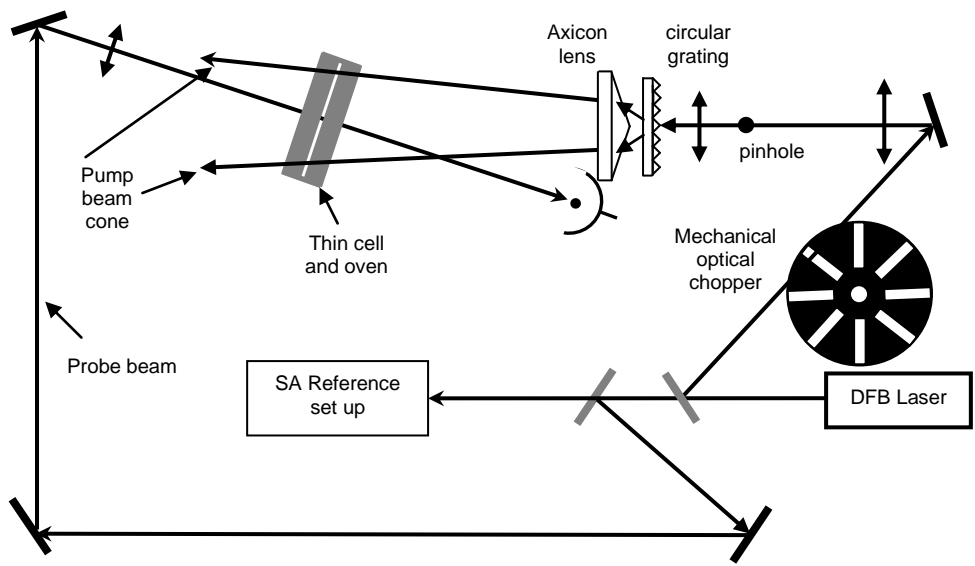

Fig. 5. *Experimental set-up*

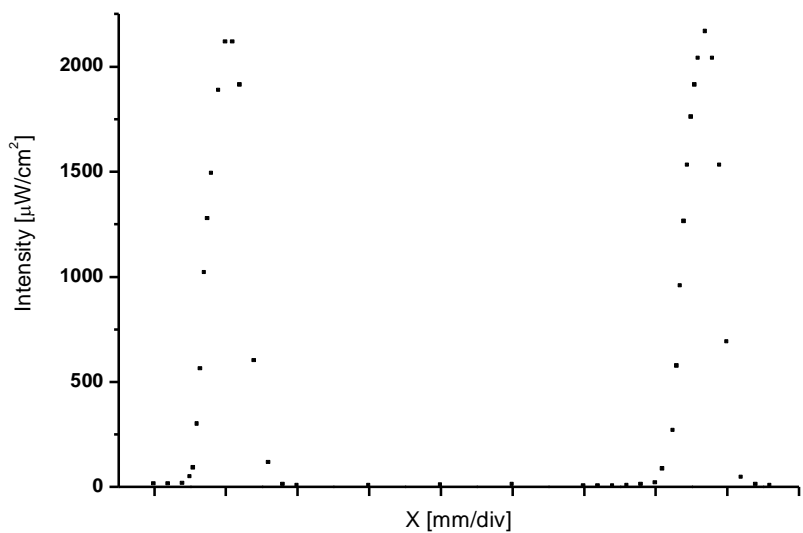

Fig. 6  *Local intensity of the ring-shaped pump, as measured across a diameter with a resolution ~0.25 mm. The intensity in the central zone is at least 200 times lower than in the regions of the ring*

We are starting to collect preliminary results, which show the possibility to observe the slow atoms with the set-up described above. Figure 7 shows a spectrum of the probe beam transmission across the Cs $D_2$ line, starting from the ground state $F_g=4$, for a pump-probe spatial separation ~1.5 mm. One distinctly observes the sub-Doppler contribution of the separate hyperfine lines. We note that the $F_g=4$ - $F_e=4$ transition appears to be the larger one, while for a linear probe spectrum, one expects the $F_g=4$ - $F_e=5$ transition to be larger than the $F_g=4$ - $F_e=4$. Actually, when using only a single laser, the efficiency of the pump beam on the non-cycling $F_g=4$ - $F_e=5$ transition is weaker than for the $F_g=$ - $F_e=4$ transition (and $F_g=4$ - $F_e=3$ transition). One also notes that there is a residual Doppler background, probably to be attributed to atoms which has collided the wall without losing their excitation received from the pump beam. Another notable point appearing on fig. 7 is the observation of a weak quadrature signal [16], that singles out the stronger and narrow contribution of the $F_g=4$ - $F_e=4$ transition. For our pump-probe separation ( ~ 1.5 mm) the atomic time-of-flight between the pump and the probe can be estimated to ~ 7.5 µs, a value in agreement with the phase shift of the signal (as can be measured in fig. 7) with respect to the 2 kHz pump modulation.

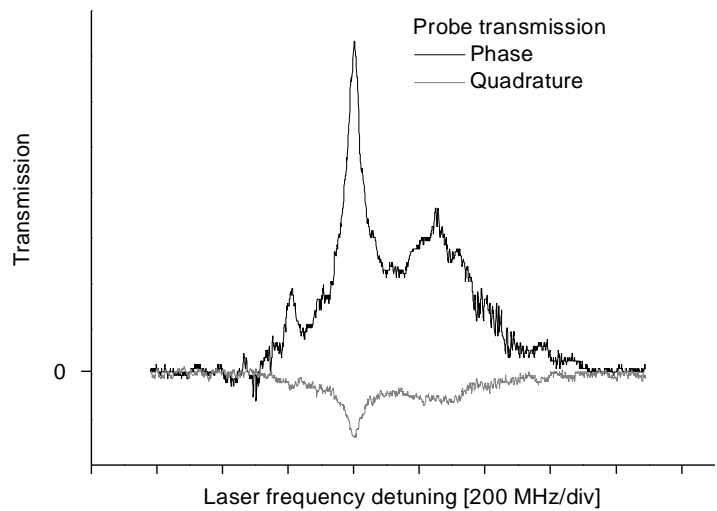

Fig. 7. *Probe beam transmission spectrum across the Cs $D_2$ line from $F_g=4$ to(left-to-right) $F_e=3,4,5$. The pump-probe separation is ~ 1.5 mm. The temperature of Cs source is 60 ºC.*

## 4. CONCLUSION

In conclusion, we are now able to observe the response of atoms flying nearly parallel to the wall, and the level of angular selectivity obtained mechanically in fig.7 (~ 1.5 mm over a 19 µm thickness) corresponds to the observation of atoms with a normal velocity ~3 m/s, smaller than the one just expected from an optical selection. It is already very comparable to the one reported [10] in an experiment where the cell thickness had not been precisely controlled. A first evaluation of the velocity distribution of the atoms leaving the surface appear now quite feasible: systematic measurements are needed for various pump distances, and the influence of residual atom-atom and atom-surface collisions must be investigated, notably by varying the Cs density, and perhaps by similar measurements in cells made with similar windows, but of different thicknesses.

## ACKNOWLEDGEMENTS

The work is performed in the frame of BNSF grant: DMU 02/17. We thank D. Sarkisyan for the preparation of the cell. The French - Bulgarian cooperation also receives a specific support from Université Paris13